\documentclass[sn-mathphys,Numbered]{sn-jnl}

\usepackage{graphicx}%
\usepackage{multirow}%
\usepackage{amsmath,amssymb,amsfonts}%
\usepackage{amsthm}%
\usepackage{mathrsfs}%
\usepackage[title]{appendix}%
\usepackage{xcolor}%
\usepackage{textcomp}%
\usepackage{manyfoot}%
\usepackage{booktabs}%
\usepackage{algorithm}%
\usepackage{algorithmicx}%
\usepackage{algpseudocode}%
\usepackage{listings}%

\begin{document}

\title{Fractional dynamics and recurrence analysis in cancer model}

\author*[1]{\fnm{Enrique C.} \sur{Gabrick}}\email{ecgabrick@gmail.com}
\equalcont{These authors contributed equally to this work.}

\author[1]{\fnm{Matheus R.} \sur{Sales}}
\equalcont{These authors contributed equally to this work.}
\author[1]{\fnm{Elaheh} \sur{Sayari}}
\equalcont{These authors contributed equally to this work.}
\author[2]{\fnm{Jos\'e} \sur{Trobia}}
\equalcont{These authors contributed equally to this work.}
\author[1,3]{\fnm{Ervin K.} \sur{Lenzi}}
\equalcont{These authors contributed equally to this work.}
\author[4]{\fnm{Fernando da S.} \sur{Borges}}
\equalcont{These authors contributed equally to this work.}
\author[1,2]{\fnm{Jos\'e D.} \sur{Szezech Jr.}}
\equalcont{These authors contributed equally to this work.}
\author[1,5,6]{\fnm{Kelly C.} \sur{Iarosz}}
\equalcont{These authors contributed equally to this work.}
\author[6,7]{\fnm{Ricardo L.} \sur{Viana}}
\equalcont{These authors contributed equally to this work.}
\author[7]{\fnm{Iber\^e L.} \sur{Caldas}}
\equalcont{These authors contributed equally to this work.}
\author[1,2,7]{\fnm{Antonio M.} \sur{Batista}}
\equalcont{These authors contributed equally to this work.}

\affil[1]{Graduate Program in Science, State University of Ponta Grossa,
84030-900, Ponta Grossa, PR, Brazil}
\affil[2]{Department of Mathematics and Statistics, State University of Ponta
Grossa, 84030-900, Ponta Grossa, PR, Brazil}
\affil[3]{Department of Physics, State University of Ponta Grossa, 84030-900,
Ponta Grossa, PR, Brazil}
\affil[4]{Department of Physiology and Pharmacology, State University of New
York Downstate Health Sciences University, 11203, Brooklyn, NY, USA}
\affil[5]{University Center UNIFATEB, 84266-010, Tel\^emaco Borba, PR, Brazil}
\affil[6]{Institute of Physics, University of S\~ao Paulo, 05508-090, S\~ao
Paulo, SP, Brazil}
\affil[7]{Department of Physics, Federal University of Paran\'a, 82590-300,
Curitiba, PR, Brazil}

\abstract{
In this work, we analyze the effects of fractional derivatives in the chaotic 
dynamics of a cancer model. We begin by studying the dynamics of a standard 
model, {\it i.e.}, with integer derivatives. We study the dynamical 
behavior by means of the bifurcation diagram, Lyapunov exponents, and
recurrence quantification analysis (RQA), such as the recurrence rate (RR), the
determinism (DET), and the recurrence time entropy (RTE). We find a high
correlation coefficient between the Lyapunov exponents and RTE. Our simulations 
suggest that the tumor growth parameter ($\rho_1$) is associated with a 
chaotic regime. Our results suggest a high correlation between the largest Lyapunov
exponents and RTE. After understanding the dynamics of the model in the standard
formulation, we extend our results by considering fractional operators. We fix
the parameters in the chaotic regime and investigate the effects of the 
fractional order. We demonstrate how fractional dynamics can be properly
characterized using RQA measures, which offer the advantage of
not requiring knowledge of the fractional Jacobian matrix. We find that the
chaotic motion is suppressed as $\alpha$ decreases, and the system becomes
periodic for $\alpha \lessapprox 0.9966$. We observe limit cycles for
$\alpha \in (0.9966,0.899)$ and fixed points for $\alpha<0.899$. The fixed point
is determined analytically for the considered parameters. Finally, we discover
that these dynamics are separated by an exponential relationship between
$\alpha$ and $\rho_1$. Also, the transition depends on a supper transient which
obeys the same relationship.
}


\maketitle

%%%%%%%%%%%%%%%%%%%%%%%%%%%%%%%%%%%%%%%%%%
%%%%%%%%%%%%%%%%%%%%%%%%%%%%%%%%%%%%%%%%%%

\section{Introduction}

Cancer is a set of diseases that arises from the abnormal growth and 
uncontrolled division of body cells. It can spread to the body's cells, 
causing many deaths \cite{Hausman2019}. Each year the American Cancer 
Society estimates the number of new cancer cases. For 2020 there was an 
estimated number of 19.3 million new cancer cases and almost 10.0 million 
deaths from cancer \cite{Ferlay2021}. Only in the United States of America 
were projected 1,918,030 new cancer cases and 609,360 cancer deaths for 2022
\cite{Siegel2022}. In this way, cancer is a crucial public health problem
worldwide \cite{Siegel2018} that requires many efforts to understand the
mechanism behind the illness and improve the treatment methods
\cite{Jones2007, Mufudza2012, Waldman2020}. There are several methods to study
the dynamics of cancer cell proliferation and one of the most successful method
is through mathematical models.

Mathematical model is a powerful tool for understanding cancer dynamics 
\cite{Lopez2019a, Iarosz2015} and simulate treatment measures, such as effects 
of drug resistance \cite{Trobia2020, Trobia2021}, immunotherapy
\cite{Castiglione2007}, chemotherapy \cite{Lopez2019b, Borges2014}, radiotherapy
\cite{Liu2014}, biochemotherapy \cite{Mamat2013} and predict fluctuations
associated with the growth rate of cancer cells \cite{Sayari2022}. A fundamental question that arises is how to understand the dynamics among healthy cells, 
the immune system and the tumor cells. In light of this question, several models
have been proposed and studied \cite{Pillis2003, Iarosz2011, Kuznetsov1994, Pinho2002, Amatruda2002, Altrock2015, Tuveson2019}.

Dehingia \textit{et al.} \cite{Dehingia2021} studied the effect of time 
delay in tumor-immune interaction and stimulation process. They obtained 
conditions for the existence of equilibrium points and Hopf bifurcation. 
Also, they derived conditions for periodic solutions. D\'iaz-Mar\'in 
\textit{et al.}~\cite{Diaz-Marin2022} proposed a model to describe cell 
population dynamics in a tumor with periodic radiation as treatment. As 
global attractors, they found almost periodic solutions or the vanishing 
equilibrium. L\'opez \textit{et al.}~\cite{Lopez2019} formulated a model 
for tumor growth in the presence of cytotoxic chemotherapeutic agents. 
Their model allows an investigation of the Norton-Simon hypothesis in 
the context of dose-dense chemotherapy. In another work, L\'opez 
\textit{et al.}~\cite{Lopez2014b} validated a model by means of experimental 
results. In this model, they considered tumor growth, including 
tumor-healthy cell interactions, immune response, and chemotherapy. Most 
of these models deal with three or more dimensional systems, where 
chaotic behavior is a possibility~\cite{Telbook}.

A three-dimensional cancer model with chaotic dynamics was proposed by 
Itik and Banks~\cite{Itik2010}. In their model, they considered the 
interactions of tumor cells with healthy host cells and immune systems. 
Through the calculation of Lyapunov exponents, they showed the existence 
of chaotic attractors for some parameters. Letellier 
\textit{et al.}~\cite{Letellier2013},  made a topological and observability 
analysis of this model. They also investigated the equilibrium points and the 
bifurcation diagrams. Khajanchi \textit{et al.}~\cite{Khajanchi2018} 
studied a similar model with time delay. They analyzed the existence 
and stability of biologically feasible points and the emergence of Hopf 
bifurcations. Chaos in the 3-cell cancer model was also studied by 
Abernethy and Gooding~\cite{Abernethy2018}, Gallas 
\textit{et al.}~\cite{Gallas2014}, Khajanchi~\cite{Khajanchi2015} and 
others~\cite{Kemwoue2020, Li2021, Valle2018, Uthamacumaran2021}. 

Although the literature about the model proposed by Itik and Banks
\cite{Itik2010} is extensive, the works are restricted to integer-order
differential equations. However, this formulation does not incorporate non-local
effects. To do that, it is necessary to consider fractional differential 
equations~\cite{Lenzi2023}. In general, non-local operators are more 
efficient in describing some situations since they capture non-localities 
and have memory effects~\cite{Lenzi2018}. {Fractional operators have
been used to model many real problems, such as} photo acoustic \cite{Somer2023, Somer2022}, viscoelastic properties~\cite{Carmo2023}, Quantum Mechanics~\cite{Cius2022, Lenzi2013, GabrickQM2023}, Epidemiology~\cite{Srivastava2020, Dong2020},
Ecology~\cite{Kumar2022, Mahmoud2020}, Duffing oscillator~\cite{Coccolo2023} and
many others~\cite{Nadeem2022, Vivekanandhan2023, Sun2018}.

The dynamical behavior of the cancer model proposed by Itik and Banks
\cite{Itik2010} was studied in light of fractional operators in Ref.
\cite{Atangana2018}. It was considered three different fractional derivatives
formulations: the power-law \cite{Caputo2021} (singular kernel), the exponential
\cite{Fabrizio2015}, and the Mittag-Leffler \cite{Baleanu2016}. The results show
that the dynamics is changed as a consequence of the extension to fractional 
differential operators. Ghanbari~\cite{Ghanbari2020} also explored the 
same model and investigated the influence of power-law, exponential 
decay-law, and Mittag-Leffler in the fractal-fractional approach. These 
researchers obtained conditions for the existence and uniqueness of the 
solutions and expanded a numerical method to study the dynamical behavior. 
The results showed that the dynamics change
from chaotic to a limit cycle depending on the factional order. Naik \textit{et al.} \cite{Naik2021}, considering an extension given by the Caputo derivative, analyzed the stability of the
model. In a numerical scheme, they reported that the system goes to a limit
cycle in a chaotic regime for the standard model. Xuan \textit{et al.}
\cite{Xuan2022}, using the Caputo fractal-fractional derivative in the cancer
model, {demonstrated that hidden attractors emerge under fractional
operators.} To study the system's complexity, they studied bifurcation diagrams
and stability. Their findings also showed that the system converges to a 
limit cycle attractor when the fractional order is decreased. These works
extend the model of Itik and Banks~\cite{Itik2010}. Extensions of other cancer
models can be found in Refs. \cite{Ahmad2021, Ahmed2012, Arfan2021, Baleanu2019, Debbouche2022, Hassani2021, Iyiola2014, Kumar2020, Rehaman2022, SolisPerez2019}.

In this work, we analyze the influence of fractional operators in a cancer 
model \cite{Itik2010}. As a definition of the fractional operator, we 
follow the Caputo scheme \cite{Lenzi2018}, with an order equal to $\alpha$. 
We consider the parameters in which the trajectories are chaotic
\cite{Letellier2013}. As a novelty, we characterize the model using recurrence
quantification, such as recurrence rate (RR), recurrence time entropy (RTE), and
determinism (DET). Furthermore, we localize the $\alpha$ value in which the
system transits from chaos to limit cycle utilizing RTE. Also, considering the
mean square deviation of the time series, we construct the plane parameters
composed of $\rho_1 \times \alpha$, where $\rho_1$ is the tumor growth. Our
results suggest an exponential relation between $\alpha$ and tumor growth rate
$\rho_1$ in which the dynamics transit from the limit cycle to a fixed point.
This transition depends on a supper transient time that follows the same
relationship between $\rho_1$ and $\alpha$.

Our work is organized as follows. In Section \ref{sec:standard_model}, 
we revisit the integer-order model analyzing the fixed points and their 
stabilities. Section \ref{sec:fractional_approach} discusses the fractional 
influences in the cancer model. Our conclusions are drawn in Section
\ref{sec:conclusions}. 

%%%%%%%%%%%%%%%%%%%%%%%%%%%%%%%%%
%%%%%%%%%%%%%%%%%%%%%%%%%%%%%%%%%

\section{Standard model}
\label{sec:standard_model}

We consider a cancer model that describes the interactions among host 
cells ($H$), effector immune cells ($E$) and tumor cells ($T$), governed 
by the following equations:
\begin{eqnarray}
\frac{dH}{dt} &=& \rho_1 H \left(1 - \frac{H}{\kappa_1}\right) - \alpha_{13}TH, \label{eqa} \\
\frac{dE}{dt} &=& \rho_2\frac{ TE}{T + \kappa_2} -\alpha_{23} TE - \delta_2  E, \label{eqb} \\
\frac{dT}{dt} &=& \rho_3 T \left(1 - \frac{T}{\kappa_3}\right) - \alpha_{31}TH - \alpha_{32}TE. \label{eqc}
\end{eqnarray}
Equation \ref{eqa} describes the growth of the host cells by a logistic 
function with a rate equal to $\rho_1$ and biotic capacity equal to $\kappa_1$. 
The host cells are killed by tumor cells. This term is represented by the 
interaction $-\alpha_{13}TH$, where $\alpha_{13}$ is the host cell killing 
rate by tumor cells. Equation \ref{eqb} gives the rate at which immune cells 
appear. These cells are the tumor antigens whose growth depends on the quantity
of $T$, the growth rate of effector immune cells ($\rho_2$), and a positive
constant $\kappa_2$. The tumor cell not only contributes to the growth of immune
cells but also to its death. In the term $-\alpha_{23}TE$, $\alpha_{23}$ is the
rate at which tumor cells inhibit the immune cells. In addition, the $E$ cells
die according to $\delta_2$. The last equation, Eq.~\ref{eqc}, gives the
dynamics of the tumor cells. Similarly, with the $H$ cells, the tumor growth is
given by a logistic function with a growth rate equal to $\rho_3$ and biotic
capacity equal to $\kappa_3$. The $T$ cells are killed by $H$ cells at a rate
equal to $\alpha_{31}$ and by the $E$ cells with a rate equal to $\alpha_{32}$.

\begin{figure}[hbt]
\centering
\includegraphics[scale=0.30]{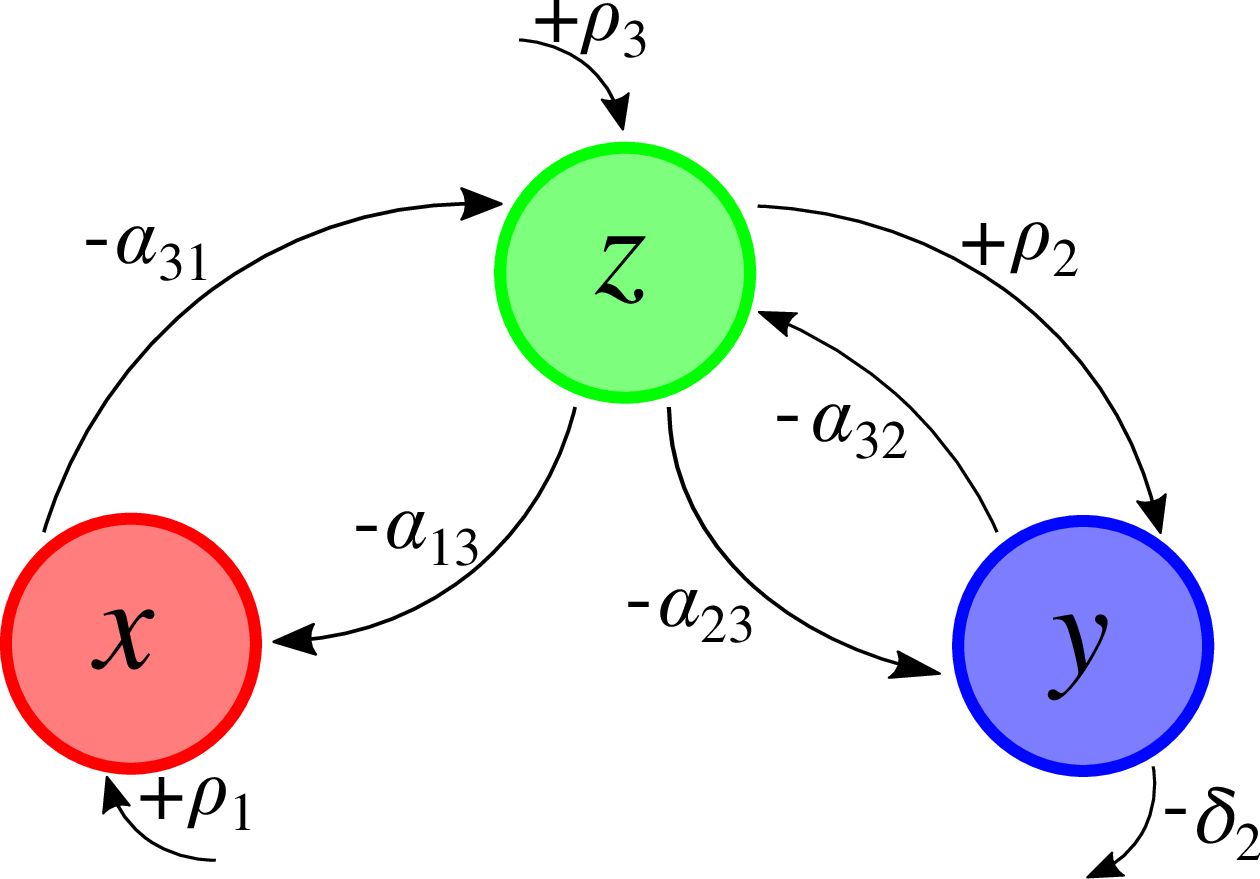}
\caption{Schematic representation of the cancer model. $x$, $y$ and $z$ 
represent the host, immune, and tumor cells. $\rho_1$ is the growth rate of $x$,
$\alpha_{13}$ is the death rate of $x$ due to $z$, $\alpha_{31}$ is the death
rate of $z$ due to $x$, $\rho_3$ is the growth rate of $z$, $\alpha_{32}$ is the
death rate of $z$ due to $y$, $\rho_2$ is the growth rate of $y$ due to $z$,
$\alpha_{32}$ is the death rate of $y$ due to $z$ and $\delta_2$ is the natural
death of $y$.}
\label{fig1}
\end{figure}

Considering the transformation $(H,E,T) \rightarrow (x,y,z)$ \cite{Itik2010},
the normalized equations are
\begin{eqnarray}
\frac{dx}{dt} &=& \rho_1 x(1-x) - \alpha_{13}xz, \label{eq1}\\
\frac{dy}{dt} &=& \frac{\rho_2 yz}{1+z} - \alpha_{23}yz - \delta_2 y, \label{eq2}
\\
\frac{dz}{dt} &=& z(1-z) - xz - \alpha_{32}yz \label{eq3}.
\end{eqnarray}
A schematic representation of the model is displayed in Fig. \ref{fig1}. The
red, green, and blue circles represent the $x$, $y$, and $z$ cells, respectively.
The arrows indicate the interactions. If the interaction is constructive (growth
cells), we denote with a signal $+$. If the interaction is destructive (kill
cells), we denote it by a signal $-$. The $x$ cells growth with a rate $\rho_1$,
which is indicated by $+\rho_1$ in Fig.~\ref{fig1}. The $x$ cells are killed by
the tumor cells at a rate equal to $\alpha_{13}$. The tumor cells grow according
to a rate $\rho_3$ and are killed by $x$ cells at a rate $\alpha_{31}$ and by
$y$ at a rate equal to $\alpha_{32}$. The $y$ cells grow when interacting with
tumor cells by a rate $\rho_2$ and are killed by $z$ cells according to a rate
equal to $\alpha_{23}$. Also, the $y$ cells have a natural death $\delta_2$. The
immune cells do not interact with the host cells. In this way, the tumor cells
are the generalist competitors and the other two are specialist competitors
\cite{Letellier2013}. In this type of interaction, there is no particular prey.

The fixed point solutions of this model are found by solving:
\begin{eqnarray}
0 &=& \rho_1 x(1-x) - \alpha_{13}xz, \label{eq4} \\
0 &=& \frac{\rho_2 yz}{1+z} - \alpha_{23}yz - \delta_2 y, \label{eq5}\\
0 &=& z(1-z) - xz - \alpha_{32}yz, \label{eq6}
\end{eqnarray}
admitting seven solutions, that are $F_i \equiv (x^i,y^i,z^i)$, where
$i=1,...,7$. Solving Eqs.~\ref{eq4}, \ref{eq5} and \ref{eq6}, we obtain
%\begin{widetext}
\begin{eqnarray}
F_1 &=& (0,0,0), \
F_2 = (0,0,1),  \
F_3 = (1,0,0) \label{sol3} \\
%%%%%%%%%%%
F_4 &=& \Bigg(0,
\frac{\sqrt{\left(\alpha _{23}+\delta _2-\rho _2\right){}^2-4 \alpha _{23} \delta _2}+
3 \alpha _{23}+\delta _2-\rho _2}{2 \alpha _{23} \alpha _{32}}, \nonumber \\
&-&\frac{\sqrt{\left(\alpha _{23}+\delta _2-\rho _2\right){}^2-4 \alpha _{23} \delta _2}+
\alpha _{23}+\delta _2-\rho _2}{2 \alpha _{23}}\Bigg) \label{sol4} \\
%%%%%%%%%%%
F_5 &=& \Bigg(\frac{\alpha _{13} \left(\sqrt{-2 \rho _2 \left(\alpha _{23}+\delta _2\right)+
\left(\alpha _{23}-\delta _2\right){}^2+\rho _2^2}+\alpha _{23}+\delta
   _2-\rho _2\right)}{2 \alpha _{23} \rho _1}+1, \nonumber \\
&-&\frac{\left(\alpha _{13}-\rho _1\right) \left(\sqrt{\left(\alpha _{23}+\delta _2-\rho _2\right){}^2-
4 \alpha _{23} \delta _2}+\alpha _{23}+\delta
   _2-\rho _2\right)}{2 \alpha _{23} \alpha _{32} \rho _1} , \nonumber \\
&-&\frac{\sqrt{\left(\alpha _{23}+\delta _2-\rho _2\right){}^2-4 \alpha _{23} \delta _2}+\alpha _{23}+
\delta _2-\rho _2}{2 \alpha _{23}}\Bigg), \label{sol5} \\
%%%%%%%%%%%%%
F_6 &=& \Bigg(0,-\frac{\sqrt{\left(\alpha _{23}+\delta _2-
\rho _2\right){}^2-4 \alpha _{23} \delta _2}-3 \alpha _{23}-\delta _2+\rho _2}{2 \alpha _{23} \alpha _{32}}, \nonumber \\
&&\frac{\sqrt{\left(\alpha _{23}+\delta _2-\rho _2\right){}^2-4 \alpha _{23} \delta _2}-
\alpha _{23}-\delta _2+\rho _2}{2 \alpha _{23}}\Bigg), \\
%%%%%%%%%%%%%%%%%
F_7 &=& \Bigg(\frac{\alpha _{13} \left(-\sqrt{-2 \rho _2 \left(\alpha _{23}+\delta _2\right)+\left(\alpha _{23}-\delta _2\right){}^2+\rho _2^2}+\alpha _{23}+\delta
   _2-\rho _2\right)}{2 \alpha _{23} \rho _1}+1, \nonumber \\
&& \frac{\left(\alpha _{13}-\rho _1\right) \left(\sqrt{\left(\alpha _{23}+\delta _2-\rho _2\right){}^2-4 \alpha _{23} \delta _2}-\alpha _{23}-\delta
   _2+\rho _2\right)}{2 \alpha _{23} \alpha _{32} \rho _1}, \nonumber \\
&& \frac{\sqrt{\left(\alpha _{23}+\delta _2-\rho _2\right){}^2-4 \alpha _{23} \delta _2}-\alpha _{23}-\delta _2+\rho _2}{2 \alpha _{23}}\Bigg).
\end{eqnarray}
%\end{widetext}

To simplify the analysis of the fixed points, the parameters are equal
to $\alpha_{13} = 1.5$, $\rho_2 = 4.5$, $\alpha_{23} = 0.2$, $\delta_2 = 0.5$ and
$\alpha_{32} = 2.5$. $\rho_1$ is maintained as our control parameter in the
entire work. For these values, the fixed points are $F_1 = (0,0,0)$,
$F_2 = (0,0,1)$, $F_3 = (1,0,0)$, $F_4 = (0,0.3469,0.1325)$,  
$F_5 = \frac{1}{\rho_1}(\rho_1-0.1987, 0.0530(1.5 -\rho_1), 0.1325\rho_1)$, 
$F_6 = (0, -7.1470,18.8675)$ and
$F_7 = \frac{1}{\rho_1}(\rho_1-28.3012, 7.5470(1.5 - \rho_1), 18.8675\rho_1)$. 
$F_1$ is a trivial solution and exhibits a population without cells. The 
eigenvalues associated with the Jacobian calculated in $F_1$ are  
$(1,-0.5,\rho_1)$. The signs depend on $\rho_1$, which is always positive. 
Then, we find two unstable and one stable point, which characterize a 
saddle. $F_2$ is the point that represents a situation where the cancer 
cells occupy the whole host. The Jacobian eigenvalues in $F_2$ are
$(1.55, \rho_1-1.5, -1)$. Considering $\rho_1 = 0.6$, we obtain
$(1.55,-0.9,-1)$, which characterizes a saddle point. $F_3$ corresponds to the
cancer-free solution. The Jacobian eigenvalues calculated at this point are
$\left(0,-0.5,-\rho _1\right)$, as $\rho_1 \geq 0$, we have two negative
eigenvalues and one equal zero. Thus, this point is a saddle. In $F_4$,
all coordinates are positives; all solutions are in the positive octant.
The Jacobian eigenvalues in this point are
$(\rho_1-0.198754, \,\, -0.0662515+0.6131239\,i,\,\, -0.0662515-0.6131239\,i)$. 
The $F_5$ point has a biological significance for $\rho_1 \in [0.1987,  1.5]$. 
Selecting $\rho_1 = 0.6$, we obtain $(0.6688, 0.0795,  0.1325)$ which is 
an unstable point. The $F_6$ point has a negative term associated with 
the effector immune cells. Therefore, this solution does not have biological 
relevance. $F_7$ has a biological relevance when $\rho_1 \geq 28.3012$. However,
considering this range, the second element of $F_7$ is negative. This way,
the point $F_7$ is irrelevant.

\begin{figure}[hbt]
\centering
\includegraphics[scale=0.35]{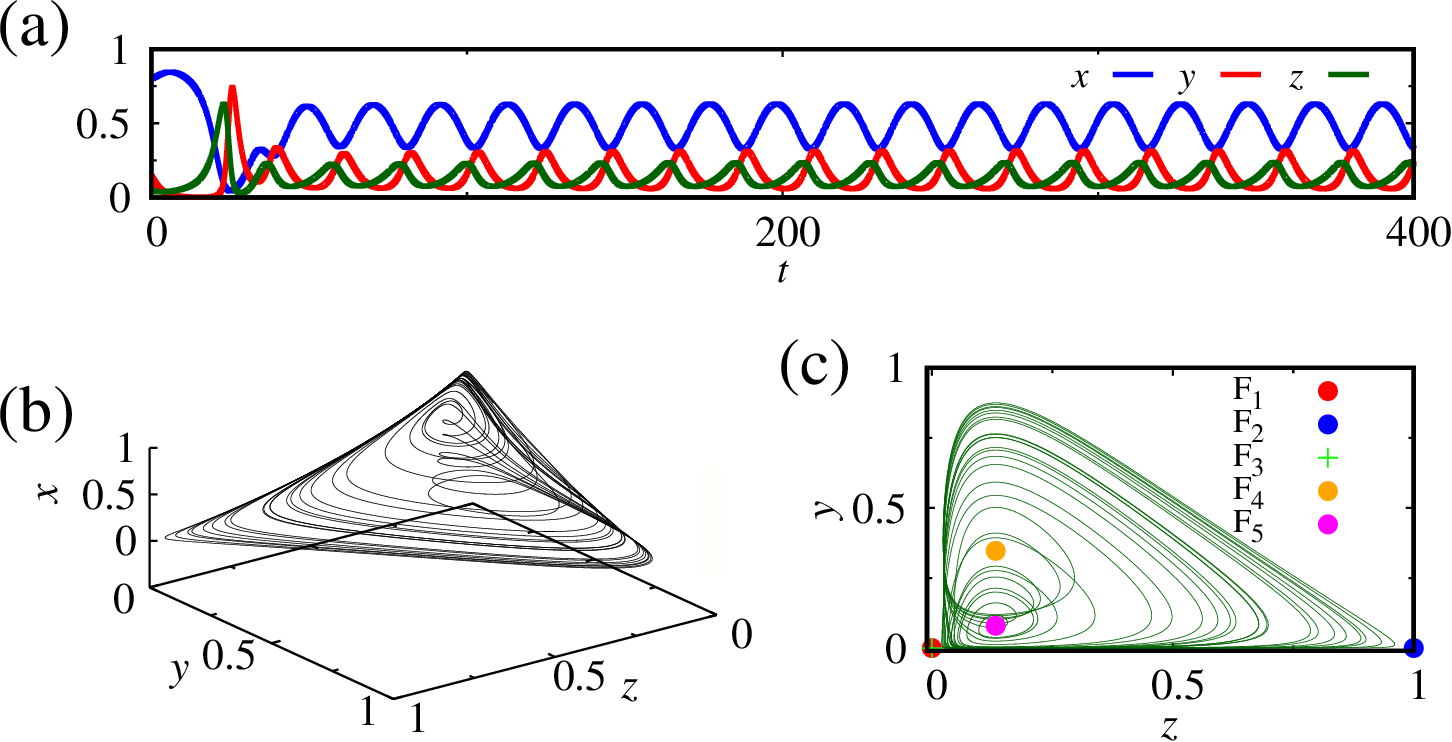}
\caption{(a) Time series for $x$, $y$ and $z$ in blue, red and green lines,
respectively. In panel (a), we consider $\rho_1 = 0.4$. (b) Phase space in
3-dimensional space and projection in the plane $y-z$ in the panel (c) for
$\rho_1 = 0.6$. We consider $\alpha_{13}=1.5$, $\rho_2=4.5$, $\alpha_{23}=0.2$,
$\delta_2=0.5$, $\alpha_{32}=2.5$, $x_0=0.80$, $y_0=0.15$, and $z_0=0.05$.}
\label{fig2}
\end{figure}

The numerical solutions for Eqs.~\ref{eq1} (blue line), \ref{eq2} (red line) and
\ref{eq3} (green line) are displayed in Fig.~\ref{fig2}(a) for $\rho_1 = 0.4$,
$\alpha_{13} = 1.5$, $\rho_2 = 4.5$, $\alpha_{23} = 0.2$, $\alpha_{32} = 2.5$ and
$\delta_2 = 0.5$. For these parameters, the solution is a limit cycle. The
solutions show that the number of host cells decreases while the immune and
cancer increase. After that, the solution reaches an oscillatory behavior
between the three population cells. For different $\rho_1$ values, it is
possible to observe chaotic solutions. As shown in previous works
\cite{Letellier2013}, the dynamical system has a chaotic attractor for
$\rho_1 = 0.6$, as exhibited in Fig. \ref{fig2}(b). Fig. \ref{fig2}(c) shows the
2-dimensional projection together with the fixed points.

In order to distinguish periodic dynamics from chaotic ones, we use the 
recurrence plots (RPs), namely, recurrence quantification analysis (RQA), which
remain valid under fractional operators. The RPs, introduced by Eckmann
\textit{et al.} \cite{eckmann}, provide a visual representation of recurrences
of the states of a system in a $d$-dimensional phase space within a small
deviation $\epsilon$, as well as a quantitative analysis of the dynamical behavior exhibited by the system. For this purpose, given a trajectory
$\vec{x}(t) \in \mathbb{R}^d$, we define a recurrence matrix $\textbf{R}$ as
\begin{equation}
    \label{eq:recmat}
    R_{ij} = H(\epsilon - \|\vec{x}(t_i) - \vec{x}(t_j)\|),
\end{equation}
where $i, j = 1, 2, \ldots, N$, $N$ is the length of the time series, 
$H$ is the Heaviside unit step function, $\epsilon$ is a small threshold 
and $\|\vec{x}(t_i) - \vec{x}(t_j)\|$ is the spatial distance between 
two states, $\vec{x}(t_i)$ and $\vec{x}(t_j)$, in phase space in terms 
of a suitable norm.

The recurrence matrix is symmetric and binary with the elements 
$0$ representing the non-recurrent states and  the elements $1$ representing 
the recurrent ones. Two states are recurrent when the state at $t = t_i$ 
is close (up to a distance $\epsilon$) to a different state at $t = t_j$. 
For the threshold $\epsilon$, we choose it to be $\epsilon = 0.01$. For 
a discussion on the choice of $\epsilon$, see  Appendix.

Several measures based on the RPs have been proposed
\cite{rqa1,rqa2,rqa3,rqa4,rqa5}. The most simple of them is the recurrence 
rate, RR, defined as
\begin{equation}
    \label{eq:rr}
    \mathrm{RR} = \frac{1}{N^2}\sum_{i,j=1}^N R_{ij},
\end{equation}
which is a measure of the density of recurrent points. There are also 
measures based on the diagonal lines formed in an RP, such as determinism 
(or predictability)
\begin{equation}
    \label{eq:det}
    \mathrm{DET} = \frac{\sum_{\ell = \ell_{min}}^{N}\ell P(\ell)}{\sum_{\ell = 1}^{N} \ell P(\ell)},
\end{equation}
where $P(\ell)$ is the total number of diagonal lines with length $\ell$ and
minimal length $\ell_{min}$. Determinism is also a
density as the recurrence rate. It is the density of recurrent points that lie on a diagonal line and
refers to the degree to which the dynamics of a system is predictable and
repeatable over time. Systems with stochastic or chaotic behaviors cause no or
very short diagonal lines, whereas deterministic behavior exhibits longer
diagonal lines. Therefore, the determinism is expected to be small in the first
case and large in the second case. There are also measures based on the vertical
lines of length $v$, such as the laminarity (LAM) and the trapping time (TT),
for example. For a detailed discussion about these and other measures, see 
Refs. \cite{rqa1,rqa2,rqa3,rqa4,rqa5} and references therein.

Entropy-based measures of RPs have been employed to characterize the dynamical behavior of nonlinear systems 
\cite{entropy1,entropy2,entropy3,entropy4,entropy5,entropy6,entropy7}. 
The entropy of the distribution of recurrence times (recurrence time entropy)
has been utilized as a tool for the detection of chaotic orbits. The vertical
distances between the diagonal lines, \textit{i.e.}, the gaps between them, are
an estimate of the recurrence times of the trajectory
\cite{wvl1,wvl2,wvl3,wvl4}. The Shannon entropy of the distribution of white
vertical lines (the estimate of the recurrence times) is defined as
\cite{entropy1,entropy7,entropy6}
\begin{equation}
    \mathrm{RTE} = -\sum_{v = v_{\mathrm{min}}}^{v = v_{\mathrm{max}}}p_w(v)\ln{p_w(v)},
\end{equation}
where $v_{\mathrm{min}}$ ($v_{\mathrm{max}}$) is the length of the shortest (longest)
white vertical line. $p_w(v) = P_w(v)/N_w$ and $P_w(v)$ are the relative
distribution and the total number of white vertical lines of length $v$,
respectively, and $N_w$ is the total number of them. Due to the finite size of a
RP, the distribution of white vertical lines might be biased by the white border
lines, which are cut short by the border of the RP, thus affecting the RQA
measures, such as the RTE \cite{border}. In order to avoid these border effects,
it is necessary to exclude the distribution of the lines that begin and end at
the border of the RP.

\begin{figure}[hbt]
\centering
\includegraphics[scale=0.34]{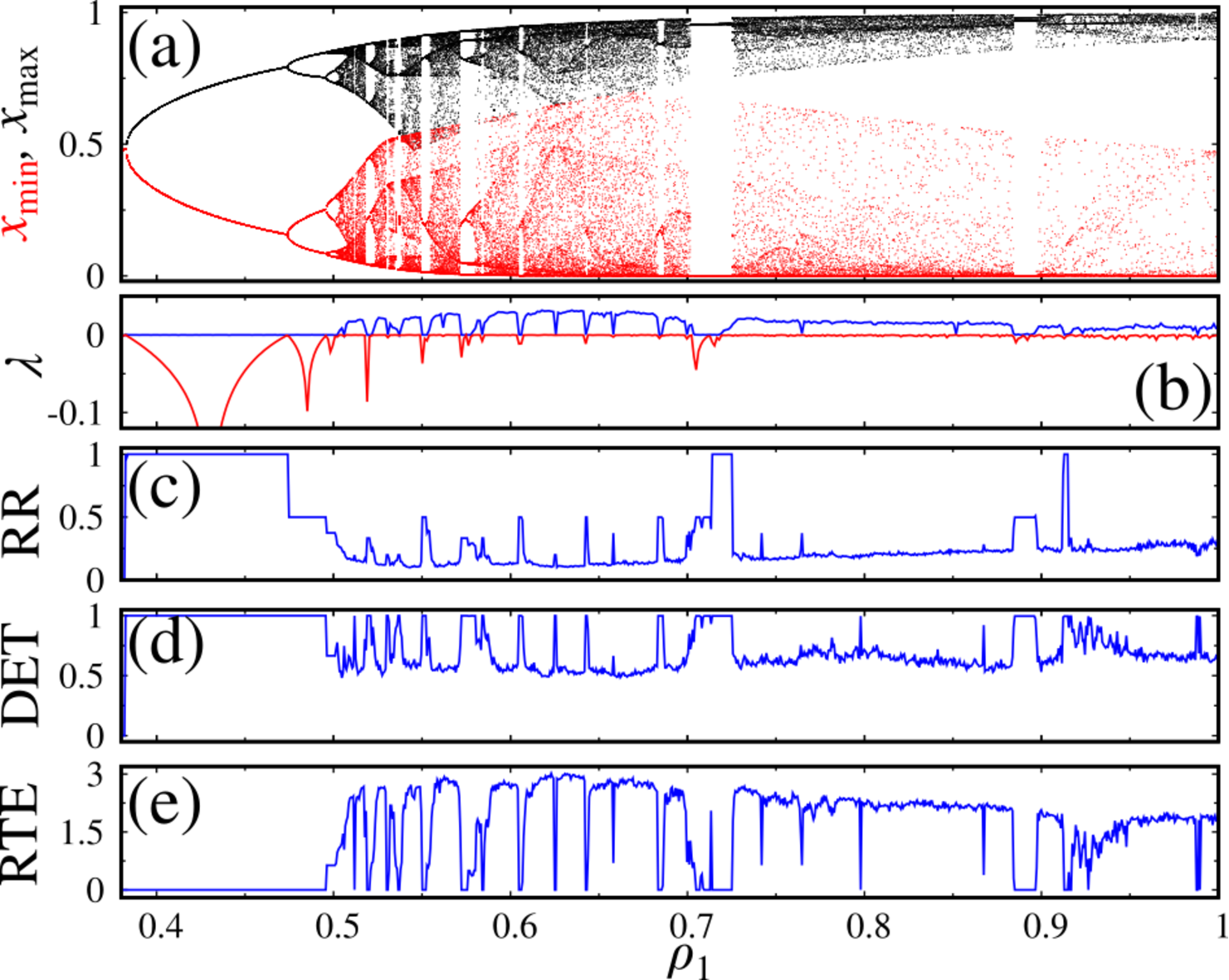}
\caption{(a) Bifurcation diagram of $x_{\rm max}$ (black points) and $x_{\rm min}$
(red points), (b) largest (blue line) and second largest Lyapunov (red line)
exponents, (c) recurrence rate, RR, (d) determinism, DET, and (e) recurrence time
entropy, RTE, as a function of $\rho_1$. We consider
$\alpha_{13} = 1.5$, $\rho_2 = 4.5$, $\alpha_{23} = 0.2$, $\delta_2 = 0.5$,
$\alpha_{32} = 2.5$, and $\epsilon = 0.01$.}
\label{fig4}
\end{figure}

To investigate the dynamic behavior of the cancer model, we consider 
$\rho_1$ as the control parameter and compute the bifurcation diagram by 
recording the local maxima, $x_{\rm max}$ (black dots) and local minima, 
$x_{\rm min}$ (red dots), of $x(t)$, as shown in Fig. \ref{fig4}(a). For 
$\rho_1<0.5$, the dynamics is periodic and becomes chaotic via period doubling.
The bifurcation diagram exhibits some periodic windows. We consider the Lyapunov
exponents ($\lambda$) as a classical method to compute chaotic solutions.
Figure \ref{fig4}(b) shows the largest, $\lambda_1$ and second largest,
$\lambda_2$, Lyapunov exponents by the blue and red lines, respectively.
The chaotic regimes are marked by at least one Lyapunov exponent greater than
zero. In the periodic windows, we observe $\lambda_1 \approx 0$ and
$\lambda_2 < 0$. On the other hand, when the dynamics is chaotic, $\lambda_1>0$.
We also calculate the RQA measures RR, DET, and RTE in panels (c), (d), and (e),
respectively. To construct the recurrence matrix, we consider the time series of
$x_{\rm max}$, $\{x_{\rm max}^{(i)}\}_{i=1,2,\ldots,N}$. We verify that periodic
solutions exhibit large values of RR and DET, with $\mathrm{DET} \rightarrow 1$,
indicating that all recurrent points lie on a diagonal line, whereas the RTE is
close to zero in these cases. RR and DET show
smaller values during the chaotic windows, while the RTE is larger. Therefore, there is a correlation
between $\lambda_1$ and these RQAs measures. To quantify this correlation, we
use the Pearson correlation coefficient, 
\begin{equation}
    \label{eq:corrcoef}
    \rho_{x, y} = \frac{{\rm cov} (x,y)}{\sigma_x\sigma_y},
\end{equation}
where ${\rm cov}(x, y)$ is the covariance between the two data series and 
$\sigma$ their respective standard deviation. We obtain for the 
correlation coefficient between $\lambda_1$ and RR, DET, and RTE, the values 
$\rho_{\lambda_1, \mathrm{RR}} = -0.54$, 
$\rho_{\lambda_1, \mathrm{DET}} = -0.41$, $\rho_{\lambda_1, \mathrm{RTE}} = 0.88$, 
respectively. Thus, the RTE is a great alternative for the characterization 
of the dynamics of this system.

%%%%%%%%%%%%%%%%%%%%%%%%%%%%%%%%%
%%%%%%%%%%%%%%%%%%%%%%%%%%%%%%%%%

\section{Fractional approach}
\label{sec:fractional_approach}

The fractional extension of the model described by Eqs.~\eqref{eq1}, 
\eqref{eq2} and \eqref{eq3} is obtained by making the following substitution:
$D f \rightarrow D^\alpha f$, where $D$ is the integer differential operator,
i.e., $df/dt$ and $\;_{0}D_{t}^\alpha f$ is the fractional differential operator
and is defined, in the Caputo sense, by
\begin{equation}
\;_{0}D_{t}^\alpha f \equiv 
\frac{1}{\Gamma\left(1-\alpha\right)}
\int_{0}^{t}dt'\frac{1}{(t-t')^{\alpha}}\frac{\partial}{\partial t}f (\vec{r},t),
\end{equation}
where $\Gamma(\cdot)$ is the gamma function and $0<\alpha<1$ \cite{Lenzi2018}. 
Therefore, the extended cancer model is given by
\begin{eqnarray}
\;_{0}D_{t}^\alpha x &=& \rho_1 x(1-x) - \alpha_{13}xz \;, \\
\;_{0}D_{t}^\alpha y &=& \frac{\rho_2 yz}{1+z} - \alpha_{23}yz - \delta_2 y\;, \\
\;_{0}D_{t}^\alpha z &=& z(1-z) - xz - \alpha_{32}yz\;.
\end{eqnarray}

\begin{figure}[hbt]
\centering
\includegraphics[scale=0.35]{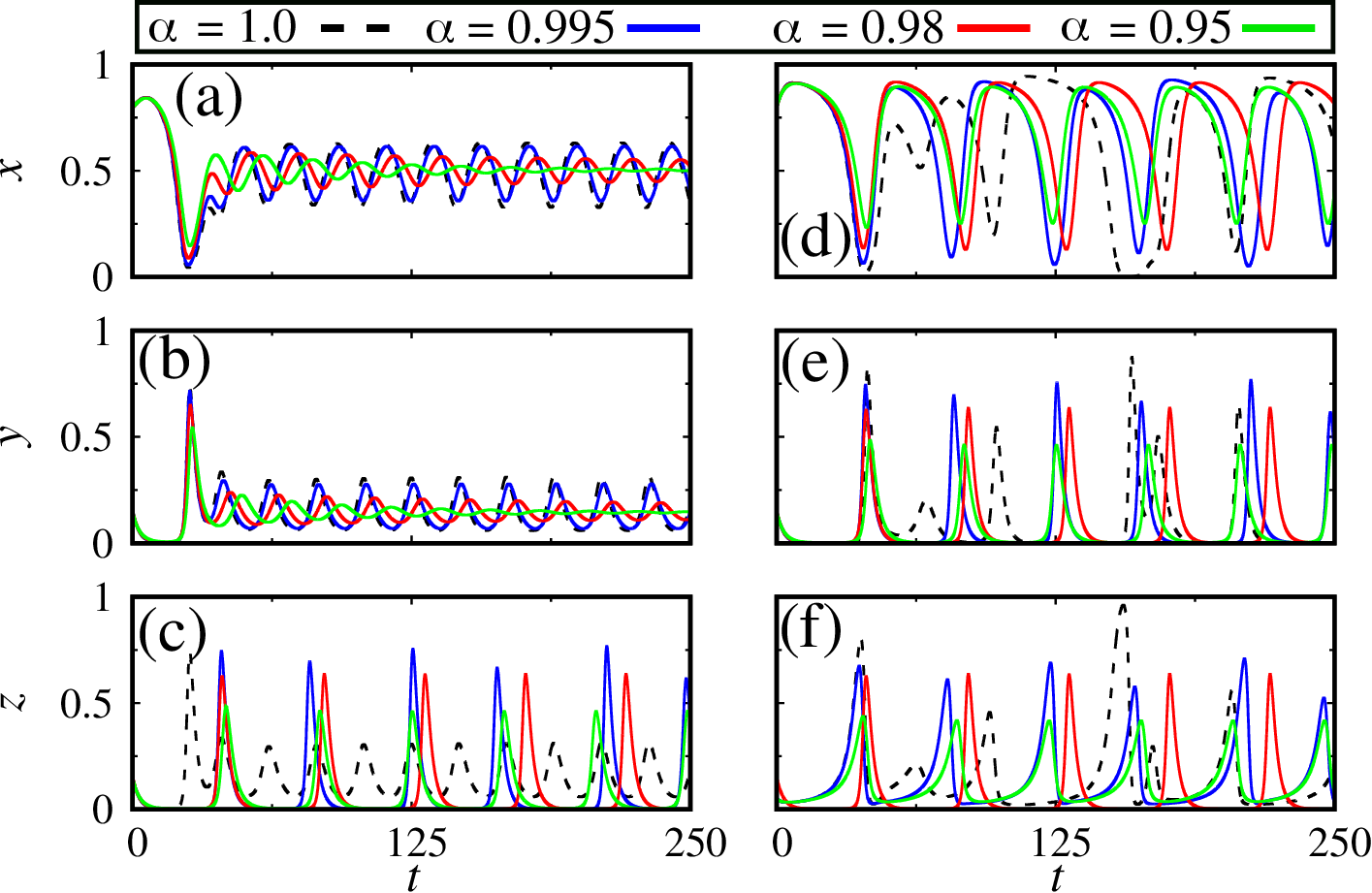}
\caption{Time series for $x$ in the panels (a) and (d), for $y$ in the panels
(b) and (e), and for $z$ in panels (c) and (f). The panels (a-c) are for
$\rho_1 = 0.4$ and (d-f) for $\rho_1 = 0.6$. The black dotted lines are for
$\alpha=1.0$, the blue line is for $\alpha=0.995$, the red line is for
$\alpha=0.98$ and the green line is for $\alpha=0.95$. We consider
$\alpha_{13} = 1.5$, $\rho_2 = 4.5$, $\alpha_{23} = 0.2$, $\delta_2 = 0.5$,
$\alpha_{32} = 2.5$, $x_0 = 0.80$, $y_0 = 0.15$, and $z_0 = 0.05$.}
\label{fig5}
\end{figure}

The time series for $\alpha = 0.995$ (blue line), $\alpha = 0.98$
(red line) and $\alpha = 0.95$ (green line) are displayed in Fig.~\ref{fig5}.
The numerical integration is made by the algorithm described in
Ref.~\cite{Garrappa2017}. The black dotted lines are for the standard case, as
exhibited in Fig.~\ref{fig2}. These results show that the fractional order
attenuates the oscillations until a fixed point as $\alpha$ decreases. For
instance, for $\alpha=0.95$, the fixed point is equal to $(x,y,z)=(0.5032, \ 0.1456, \ 0.1325)$, which corresponds to $F_5$ calculated for $\rho_1 = 0.4$.

\begin{figure}[hbt]
\centering
\includegraphics[scale=0.15]{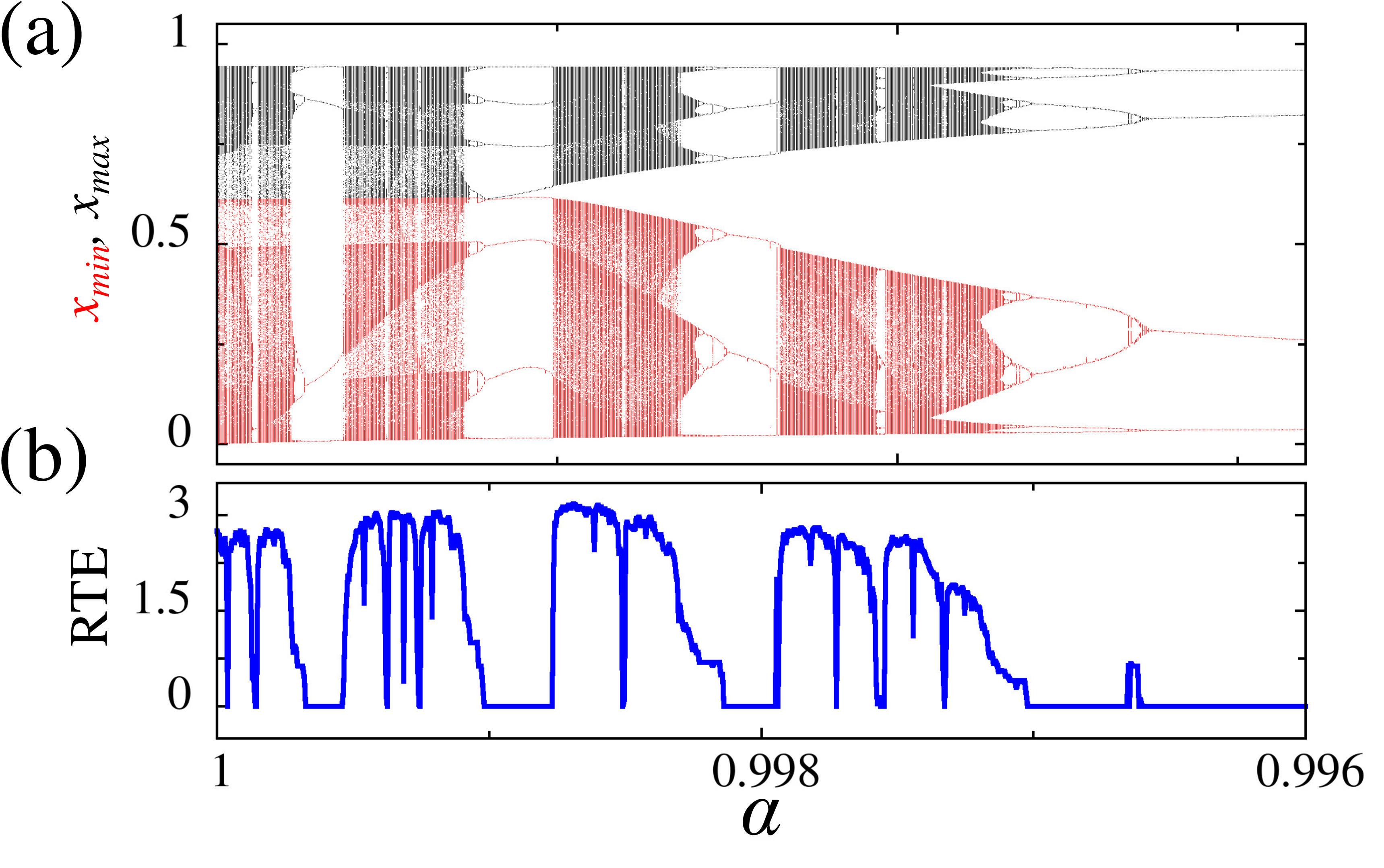}
\caption{(a) RTE and (b) bifurcation diagram as a function of $\alpha$. We
consider $\alpha_{13} = 1.5$, $\rho_1 = 0.6$, $\rho_2 = 4.5$, $\alpha_{23}=0.2$,
$\delta_2 = 0.5$, $\alpha_{32} = 2.5$, $x_0 = 0.80$, $y_0 = 0.15$, $z_0 = 0.05$,
and $\epsilon = 0.01$.}
\label{fig6}
\end{figure}

\begin{figure}[hbt]
\centering
\includegraphics[scale=0.6]{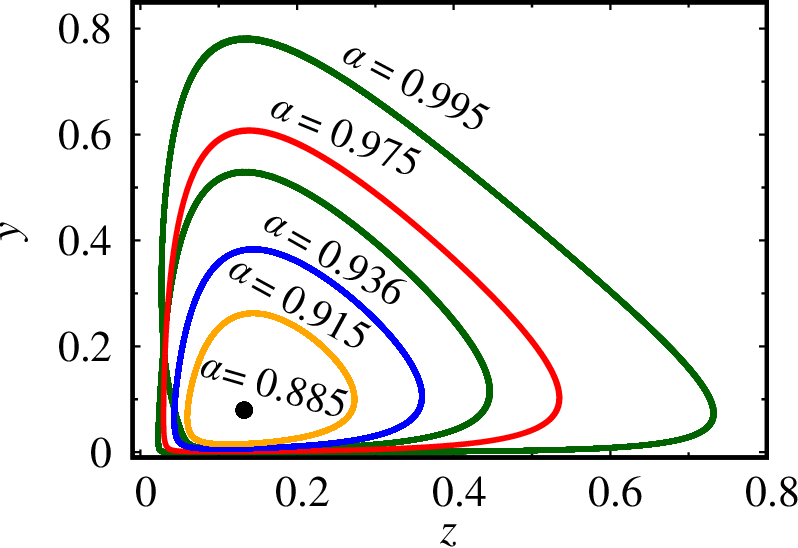}
\caption{Phase portrait for different $\alpha$ values. The green curve is for
$\alpha = 0.995$, the red curve for $\alpha = 0.975$, the blue curve for
$\alpha = 0.936$, the orange curve for $\alpha = 0.915$ and the black dot for
$\alpha = 0.885$. We consider $\rho_1 = 0.6$, $\alpha_{13} = 1.5$, $\rho_2=4.5$,
$\alpha_{23} = 0.2$, $\delta_2 = 0.5$, and $\alpha_{32} = 2.5$.}
\label{fig7}
\end{figure}

The oscillatory behavior is attenuated in the chaotic regime for $\rho_1 = 0.6$
and the dynamics is transited to periodic behavior. By computing the RTE as a
function of $\alpha$, we see that there are some periodic windows. For
$\alpha \lessapprox 0.9966$, the dynamics become periodic (Fig.~\ref{fig6}(a)).
The RTE information agrees with the bifurcation diagram, as shown in
Fig.~\ref{fig6}(b). The red and black points correspond to the $x$ minimum and
maximum values. The attractors in Fig.~\ref{fig7} display that the
periodic behavior is a limit cycle that converges to a fixed point as
$\alpha$ approaches to $0.9$. The green line is for $\alpha = 0.995$, the red
line for $\alpha = 0.975$, the blue line for $\alpha = 0.936$, the orange line
for $\alpha = 0.915$ and the black dotted point for $\alpha = 0.885$.

\begin{figure}[hbt]
\centering
\includegraphics[scale=0.6]{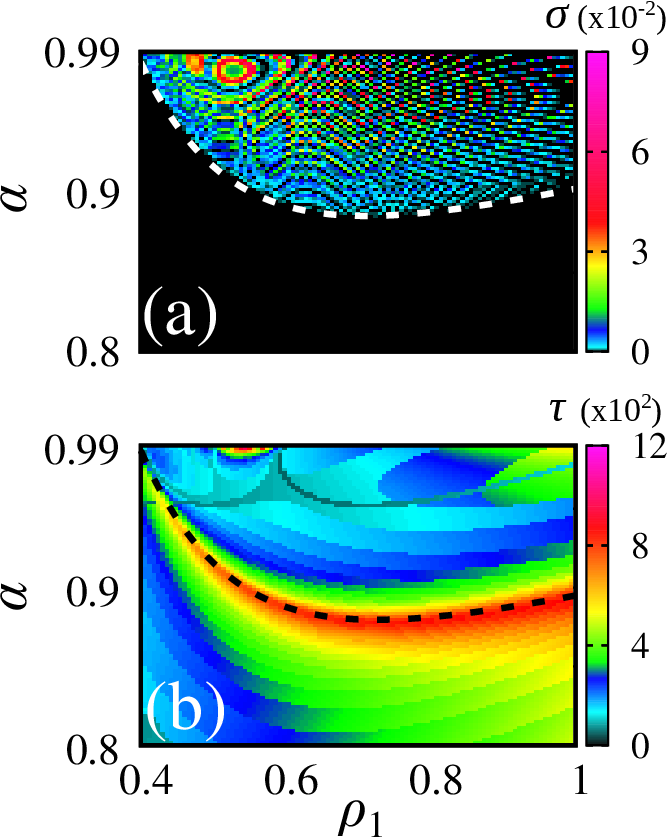}
\caption{$\rho_1 \times \alpha$ as a function of $\sigma$ (mean standard
deviation) in the panel (a) and $\tau$ (average transient time) in the panel
(b). We consider $\alpha_{13} = 1.5$, $\rho_2 = 4.5$, $\alpha_{23} = 0.2$,
$\delta_2 = 0.5$, and $\alpha_{32} = 2.5$.}
\label{fig8}
\end{figure}

One way to differentiate fixed points and limit cycles is by computing the 
mean standard deviation ($\sigma$). To do that, we discard a transient time
($\tau$). Considering this time series, we compute $\sigma$. Figure
\ref{fig8}(a) displays the parameter plane $\rho_1 \times \alpha$ as a function
of $\sigma$ in color scale. The black points indicate when the solution is a
fixed point ($\sigma < 10^{-3}$) and the colorful points indicate limit cycle
solutions. Although some black points are mixed with colorful ones, a relation between $\alpha$ and $\rho_1$ separates most parts of the
limit cycle from the fixed point solutions. This curve is indicated by the white
dotted line in the panel (a) and is given by 
$\alpha = \beta_1 e^{\beta_2 \rho_1} + \beta_3 e^{\beta_4 \rho_1}$, where
$\beta_1 = 3.150$, $\beta_2 = -7.800$, $\beta_3 = 0.808$ and $\beta_4 = 0.110$.
Figure \ref{fig8}(b) exhibits the same plane parameter with the time to the
solution for one initial condition reaches the steady solution in the color
scale. The black dotted line indicates a region of super transient, which is
defined by 
$\alpha = \xi_1 e^{\xi_2 \rho_1} + \xi_3 e^{\xi_4 \rho_1}$, where 
$\xi_1 = 3.25$, $\xi_2 = -7.7$, $\xi_3 = 0.802$, and $\xi_4 = 0.108$. Note that
$\beta_i \approx \xi_i$. In this way, the transition from the limit cycle to
fixed points crosses a supper transient region. 

\begin{figure}[hbt]
\centering
\includegraphics[scale=0.6]{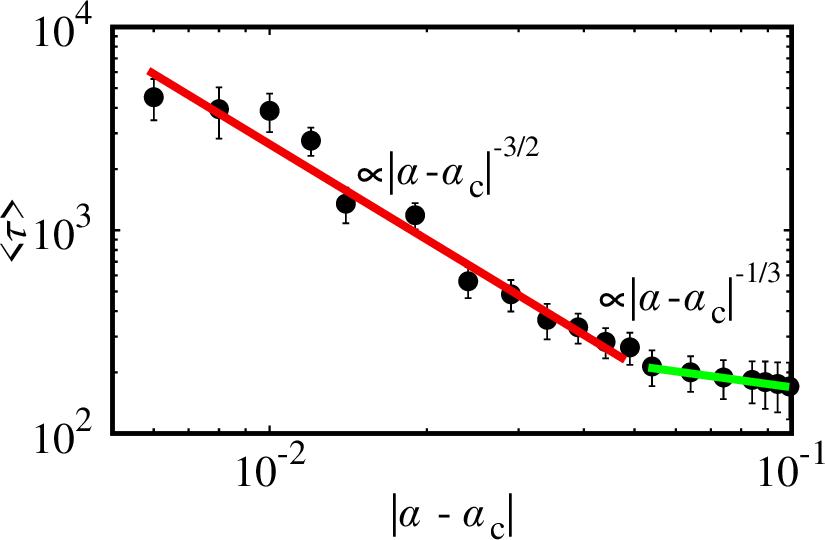}
\caption{$\langle\tau\rangle$ versus $|\alpha - \alpha_c|$, where
$\alpha_c=0.899$. The curve has two slopes, in the red curve
$\gamma \approx -3/2$ and in the green $\gamma \approx -1/3$. The respective
correlation coefficients are 0.98 and 0.99. We consider $\rho_1 = 0.6$,
$\alpha_{13} = 1.5$, $\rho_2 = 4.5$, $\alpha_{23} = 0.2$, $\delta_2 = 0.5$, and
$\alpha_{32} = 2.5$.}
\label{fig9}
\end{figure}

As our results suggest, the transition from a limit cycle to fixed points is
associated with a supper transient time. In this way, we select $\rho_1 = 0.6$
and compute the mean transient time $\langle \tau \rangle$ for 100 different
initial conditions close to the critical point. The critical point
defines the transition and equals $\alpha_c = 0.899$. Figure \ref{fig9}
shows the mean transient time ($\langle \tau \rangle$) as a function of
$|\alpha - \alpha_c|$ by the black points. The average transient lifetime scales
with $\alpha$ as a power-law according to
\begin{equation}
    \langle \tau \rangle \propto |\alpha - \alpha_c|^\gamma,
\end{equation}
and has two slopes. By fitting the data (red line in Fig. \ref{fig9}), we obtain
an exponent of $\gamma \approx -3/2$, with a correlation coefficient of 0.98 and
$\gamma \approx -1/3$ (green line) with a correlation coefficient of 0.99. All
the points evolve to $F_5$.

%%%%%%%%%%%%%%%%%%%%%%%%%%%%%%%%%%%%%%%%%%
%%%%%%%%%%%%%%%%%%%%%%%%%%%%%%%%%%%%%%%%%%

\section{Conclusions}
\label{sec:conclusions}

In this work, we consider a fractional extension of a cancer model. This 
model describes the interactions among host, effector immune, and tumor cells. Due to the non-linearity, the dynamical behavior exhibits 
chaotic solutions for some parameter ranges. For the equations that govern 
the model, we obtain analytical solutions for the fixed points and discuss 
their stability. 

In order to study a more broad case, we consider numerical solutions. As 
previously reported works, we verify chaotic solutions for $\rho_1 = 0.6$. 
Furthermore, we construct bifurcation diagrams considering the maxima and 
minimum points of the host cells. The dynamical behavior exhibited by 
the bifurcation diagram also was verified by the Lyapunov exponents and 
recurrence quantification analysis (RQA) measures. With regard to the 
RQA measures, we compute the recurrence rate (RR), determinism (DET), and 
recurrence time entropy (RTE). All three of them agree with the results 
from the Lyapunov exponents. By computing the Pearson correlation 
coefficient between them and $\lambda_1$, we observe a higher coefficient 
for the RTE. Thus, we consider only the RTE to analyze the effects of 
fractional order.  

We study the effects of fractional operators in the cancer model by the Caputo
definition. Our main goal is to understand what happens in the chaotic behavior.
By computing the RTE, our results suggest that the dynamical system becomes
periodic for $\alpha \lessapprox 0.9966$. When we look for the phase portrait,
the results show that the solutions transit from a limit cycle to a fixed point,
which is equal to the fixed points calculated analytically. The limit cycle
regime is in the range $\alpha \in (0.9966, 0.899)$ for $\rho_1 = 0.6$. For
different $\rho_1$ values, we construct the parameter space
$\rho_1 \times \alpha$ as a function of the mean standard deviation $\sigma$.
Values of $\sigma$ close to zero show fixed point solutions. Our results show
the existence of an exponential curve that delimits fixed point solutions from
limit cycles in this parameter space. Also, we analyze the transient time 
related to the transition from the limit cycle to a fixed point. For 
$\rho_1 = 0.6$, we find that for $\alpha<0.899$ the dynamics go to the fixed
point $F_5$ by a super transient with order $10^4$. The super transient curve
follows a linear decay according to two slopes, equal to $-3/2$ and $-1/3$.  

%%%%%%%%%%%%%%%%%%%%%%%%%%%%%%%%%
%%%%%%%%%%%%%%%%%%%%%%%%%%%%%%%%%

\section*{Acknowledgments}

This work was possible with partial financial support from the following 
Brazilian government agencies: CNPq, CAPES, Funda\-\c c\~ao A\-rauc\'aria 
and S\~ao Paulo Research Foundation (FAPESP 2018/03211-6, 2022/13761-9). 
R. L. V. received partial financial support from the following Brazilian 
government agencies: CNPq (403120/2021-7, 301019/2019-3), 
CAPES (88881.143103/2017-01), FAPESP (2022/04251-7).  E. K. L. received 
partial financial support from the following Brazilian government agency: 
CNPq (301715/2022-0).  E.C.G. received partial financial support from
Coordenação de Aperfeiçoamento de Pessoal de Nível Superior - Brasil (CAPES) 
- Finance Code 88881.846051/2023-01. We would like to thank the 105 Group 
Science (www.105groupscience.com).

%%%%%%%%%%%%%%%%%%%%%%%%%%%%%%%%%
%%%%%%%%%%%%%%%%%%%%%%%%%%%%%%%%%

\section*{Declarations}
{\textbf Conflict of Interest:}\\
The author declares that there exists no competing financial interest or 
personal relationships that could have appeared to influence the work 
reported in this paper.

%%%%%%%%%%%%%%%%%%%%%%%%%%%%%
%%%%%%%%%%%%%%%%%%%%%%%%%%%%%
\section*{Appendix}

\begin{figure}[hbt]
\centering
\includegraphics[scale=0.6]{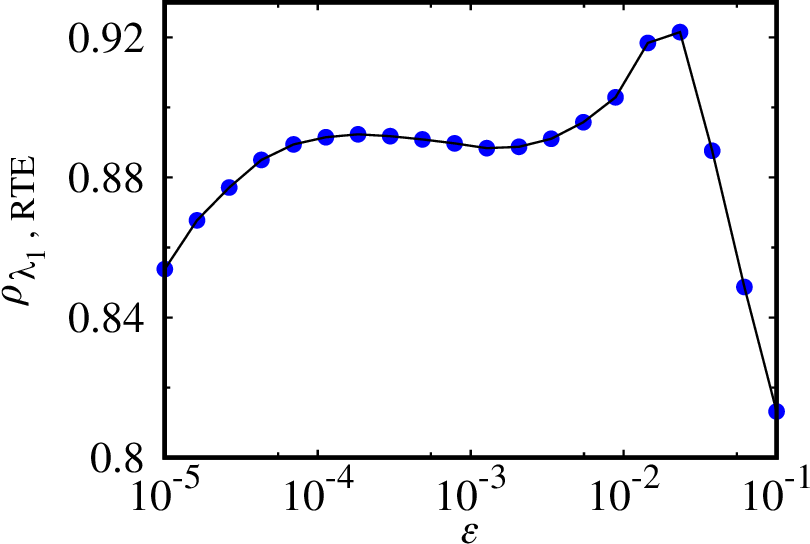}
\caption{The correlation coefficient between $\lambda_1$ and RTE as a 
function of the threshold $\epsilon$.}
\label{fig:fig10}
\end{figure}

In Section \ref{sec:standard_model}, we introduced the cancer model and 
analyzed its dynamics for different parameter values by means of a 
bifurcation diagram, the Lyapunov exponents, and three RQA measures: 
RR, DET, and RTE. We chose the threshold $\epsilon$ to be $\epsilon = 0.01$. 
Many researchers have addressed the problem of finding the best value of 
$\epsilon$ \cite{eps1,eps2,eps3,eps4}. Here, we used the correlation 
coefficient, Eq. \eqref{eq:corrcoef}, to determine $\epsilon$. We computed 
the RTE as a function of $\rho_1$ for different threshold values, ranging 
from $10^{-5}$ to $10^{-1}$, and calculated the correlation coefficient 
between $\lambda_1$ and RTE (Fig. \ref{fig:fig10}). Even if we choose a 
very small $\epsilon$ ($10^{-5}$) or a relatively large $\epsilon$ 
($10^{-1}$), we still obtain a rather high correlation coefficient, with 
the highest value being around $\epsilon \approx 10^{-2}$, and hence our 
choice of $\epsilon$. Furthermore, the fact that $\rho_{\lambda_1, \mathrm{RTE}}$ 
does not change too much within this range of $\epsilon$ indicates that, 
in our case, the choice of $\epsilon$ is not as sensible as it would be 
in other cases. In fact, there is a wide range of values for $\epsilon$ 
in which the results are good.

%%%%%%%%%%%%%%%%%%%%%%%%%%%%%%%%%%%%%%%%%%
%%%%%%%%%%%%%%%%%%%%%%%%%%%%%%%%%%%%%%%%%%

\end{document}